\UseRawInputEncoding
\RequirePackage[l2tabu, orthodox]{nag}
\documentclass[aps, prb, amsmath, twocolumn, longbibliography, superscriptaddress, footinbib, 10pt, final]{revtex4-2}
\pdfoutput=1

\usepackage{float}
\usepackage{graphicx} 
\usepackage{amssymb, amsmath}
\usepackage{blindtext}
\usepackage{siunitx} 
\usepackage[final]{microtype} 
\usepackage[T1]{fontenc}
\usepackage{lmodern}
\usepackage{xcolor}
\usepackage[colorlinks=true,linkcolor=blue, citecolor=blue, urlcolor=blue]{hyperref}
\usepackage[utf8]{inputenc}
\usepackage[acronym]{glossaries}
\usepackage{setspace}
\usepackage{graphics}

\begin{document}

\newcommand{\kp}{\bm{k.p}}  
\newcommand{\RomanNumeralCaps}[1]
    {\MakeUppercase{\romannumeral #1}}

\newcommand{\polydept}{Department of Engineering Physics, \'Ecole Polytechnique de Montr\'eal, C.P.~6079, Succ.~Centre-Ville, Montr\'eal, Qu\'ebec, Canada H3C 3A7}

\newcommand{\uToronto}{Department of Physics and Institute for Optical Sciences, University of Toronto, 60 St.~George Street, Toronto, Ontario, Canada M5S 1A7}

\newcommand{\AppliedM}{Applied Materials Inc., 974 E.~Arques Avenue, Sunnyvale, CA 94085, USA}

\newcommand{\SiGe}[2]{Si$_{#1}$Ge$_{#2}$}

\newcommand{\ac}[1]{\gls*{#1}}


\newcommand{\citesupp}{
    \footnote{See Supplemental Material at \protect \textbf  {[URL will be inserted by publisher]} for details on the other independent components of two-photon injection tensors, as well as coherent control calculations of maximal swarm velocity and circular polarizations configuration injection tensors.}}


\newacronym{sls}{SLs}{superlattices}
\newacronym{cb}{CB}{conduction band}
\newacronym{sl}{SL}{superlattice}
\newacronym{if}{IF}{interface}
\newacronym{ml}{ML}{monolayers}
\newacronym{se}{SE}{spectroscopic ellipsometry}
\newacronym{qw}{QW}{quantum well}
\newacronym{apt}{APT}{atom probe tomography}
\newacronym{hrxrd}{HRXRD}{high-resolution Xray diffraction}
\newacronym{aoi}{AOI}{angle of incidence}
\newacronym{cp}{CP}{critical point}
\newacronym{mocvd}{MOCVD}{metal-organic chemical vapor deposition}
\newacronym{haadf-stem}{HAADF-STEM}{high-angle annular dark field scanning transmission electron microscopy}
\newacronym{de}{DE}{differential evolution}
\newacronym{rms}{RMS}{surface roughness}
\newacronym{eels}{EELS}{electron energy loss spectroscopy}
\newacronym{rta}{RTA}{rapid thermal annealing}
\newacronym{afm}{AFM}{atomic force microscopy}
\newacronym{rtse}{RTSE}{room temperature spectroscopic ellipsometry}

\title{Mid-Infrared Optical Spin Injection and Coherent Control}

\author{G. Fettu}
\affiliation{\polydept{}}

\author{J. E. Sipe}
\affiliation{\uToronto{}}

\author{O. Moutanabbir}
\affiliation{\polydept{}}


\begin{abstract}

The optical injection of charge and spin currents are investigated in Ge$_{1-x}$Sn$_{x}$ semiconductors as a function of Sn content. These emerging silicon-compatible materials enable the modulation of these processes across the entire mid-infrared range. Under the independent particle approximation, the one- and two-photon interband absorption processes are elucidated, and the evolution of the coherent control is discussed for three different polarization configurations. To evaluate the contribution of high-energy transitions, a full-zone 30-band  k$\cdot$p is employed in the calculations. 
It was found that, besides the anticipated narrowing of the direct gap and the associated shift of the absorption to longer wavelengths, incorporating Sn in Ge also increases the one-photon degree of spin polarization (DSP) at the $E_1$ resonance. Moreover, as the Sn content increases, the magnitude of the response tensors near the band edge exhibits an exponential enhancement. This behavior can be attributed to the Sn incorporation-induced decrease in the carrier effective masses. This trend appears to hold also at the $E_1$ resonance for pure spin current injection, at least at low Sn compositions. The two-photon DSP at the band edge exceeds the value in Ge to reach 60 \% at a Sn content above 14 \%. These results demonstrate that Ge$_{1-x}$Sn$_{x}$ semiconductors can be exploited to achieve the quantum coherent manipulation in the molecular fingerprint region relevant to quantum sensing.
\end{abstract}
\maketitle


\UseRawInputEncoding
\section{Introduction}

\par The coherent control of quantum interference between two independent pathways is a generic process allowing the manipulation of the final state of a given system \cite{atanasov1996coherent,hache1997observation,cote1999thz,bhat2000optically,stevens2003quantum,hubner2003direct,shi2021coherence,cong2021coherent,jana2021reconfigurable,dusanowski2022optical,peng2022coherent,sederberg2022perspective}. For instance, by combining one- and two-photon absorption processes, this principle enables the injection of ballistic charge and pure spin currents in semiconductors with amplitudes and directions dependent on the incident fields' polarization and phase \cite{ruzicka2012optical,sederberg2020vectorized,jana2021reconfigurable, sederberg2022perspective}. This ability to induce relatively strong and directional charge currents stimulated further interest to implement qubit-photon interfaces needed in quantum communication \cite{dusanowski2022optical}, control phonon-photon interaction in quantum materials \cite{shi2021coherence}, and generate spintronic THz emission \cite{cong2021coherent}, and magnetic fields \cite{sederberg2020tesla, jana2021reconfigurable}. The injected currents have also been used in detection schemes to determine the parameters of the incident light beams \cite{roos2003characterization, fortier2004carrier, roos2005solid, smith2007optical} and control spectral lineshapes toward extreme ultraviolet lasing \cite{peng2022coherent}. Interestingly, current studies on semiconductors have been predominately centered on achieving optical injection in the near infrared and telecom wavelengths. This is mainly attributed to the available materials and heterostructures that are based on GaAs, Si, or Ge. Indeed,  the initial studies on the coherent control have focused on GaAs \cite{atanasov1996coherent, hache1997observation, stevens2003quantum}, followed by Si and Ge \cite{costa2007all, spasenovic2008all, loren2009optical}. Both materials have an indirect band gap, but contrary to Si, Ge allows the resonant injection of ballistic currents across its direct gap, without significant involvement of indirect transitions. Ge is also of particular interest because of its stronger spin-orbit coupling, long spin lifetime \cite{pezzoli2012optical, li2012intrinsic}, and direct gap at telecom wavelength. 

Extending the optical spin injection and coherent control to longer wavelengths would enable the quantum coherent manipulation in the molecular fingerprint range, which would open new possibilities for applications in dynamic and structural quantum sensing \cite{vitanov2019highly, xu2022coherent}. With this perspective, herein we propose Ge$_{1-x}$Sn$_{x}$ semiconductors \cite{moutanabbir2021monolithic} as a platform for a tunable coherent control over the entire mid-infrared region. As an emerging alloy, the optical, electronic, and spin properties of Ge$_{1-x}$Sn$_{x}$ remain largely unexplored. Recent studies suggested that GeSn alloys are of interest to spintronics because of their long spin lifetime \cite{de2019spin} and even larger spin-orbit coupling. Additionally, these semiconductors are compatible with silicon processing \cite{moutanabbir2021monolithic}, which makes them relevant to implement scalable on-chip mid-infrared quantum sensors and photon-spin interfaces.

In this article, by investigating alloys with a Sn content in the 0\% to 20\% range, carrier, spin, current, and spin current injection in bulk Ge$_{1-x}$Sn$_{x}$ are discussed for the two-color $\omega$ and 2$\omega$ scheme.  Note that the much weaker phonon-assisted processes are ignored to consider only the direct transitions, therefore the modification in the directness of Ge$_{1-x}$Sn$_{x}$ does not affect the optical injection results. Under the independent particle approximation, the one- and two-photon interband absorption processes are analyzed, and the results of coherent control are presented under three different polarization configurations. In order to observe high-energy features such as the $E_1$ transition, a full-zone 30-band  k$\cdot$p is employed for the calculations. The integrals in reciprocal space are performed with a linear tetrahedron method. By extending the methodology developed for GaAs and Ge \cite{rioux2010optical, rioux2012optical}, the derived equations are found to be valid in the case where there is no intermediate states at exactly the mid-level between the initial and final states of absorption. The latter condition is met above at a Sn content above 7\% of Sn, thus the presented two-photon injection and coherent control results in Ge$_{1-x}$Sn$_{x}$ in this composition range are limited at relatively lower energy. 

The outline of this paper is as follows. In Section \ref{sec:Model}, the theoretical model used to compute the optical response tensors is presented, and the Ge$_{1-x}$Sn$_{x}$ band structure properties are discussed. In Section \ref{sec:Results}, the one- and two-photon absorption processes are described, and the calculations for three different configurations of the two-color coherent control scheme are discussed. The results are summarized in Section \ref{sec:Conclusion}.


\UseRawInputEncoding
\section{Theoretical Framework}
\label{sec:Model}

This section describes the theoretical framework established to investigate the electronic structure of Ge$_{1-x}$Sn$_{x}$ semiconductors and compute the optical response tensors. In most studies on the calculations of interband optical responses in semiconductors, k$\cdot$p models with 8 and 14 bands are typically used. However, these models are valid in a restricted zone close to the $\Gamma$ point. In Ge-based semiconductors, an accurate description of the bands close to the Brillouin Zone (BZ) edges is required to capture relevant features of the band structure, such as the $E_1$ transition \cite{rioux2010optical}. Hence, in this work, a room temperature full-zone 30-band k$\cdot$p model is used for the band structure calculations based on the parameters reported recently \cite{song2019band}, as shown in Table \ref{tab:1}. To simplify the  parametrization process, the group $O^h$ model was used, neglecting any local breaking of centrosymmetry in the alloy.


\begin{figure}[h]
    \centering
    \includegraphics[scale=0.32]{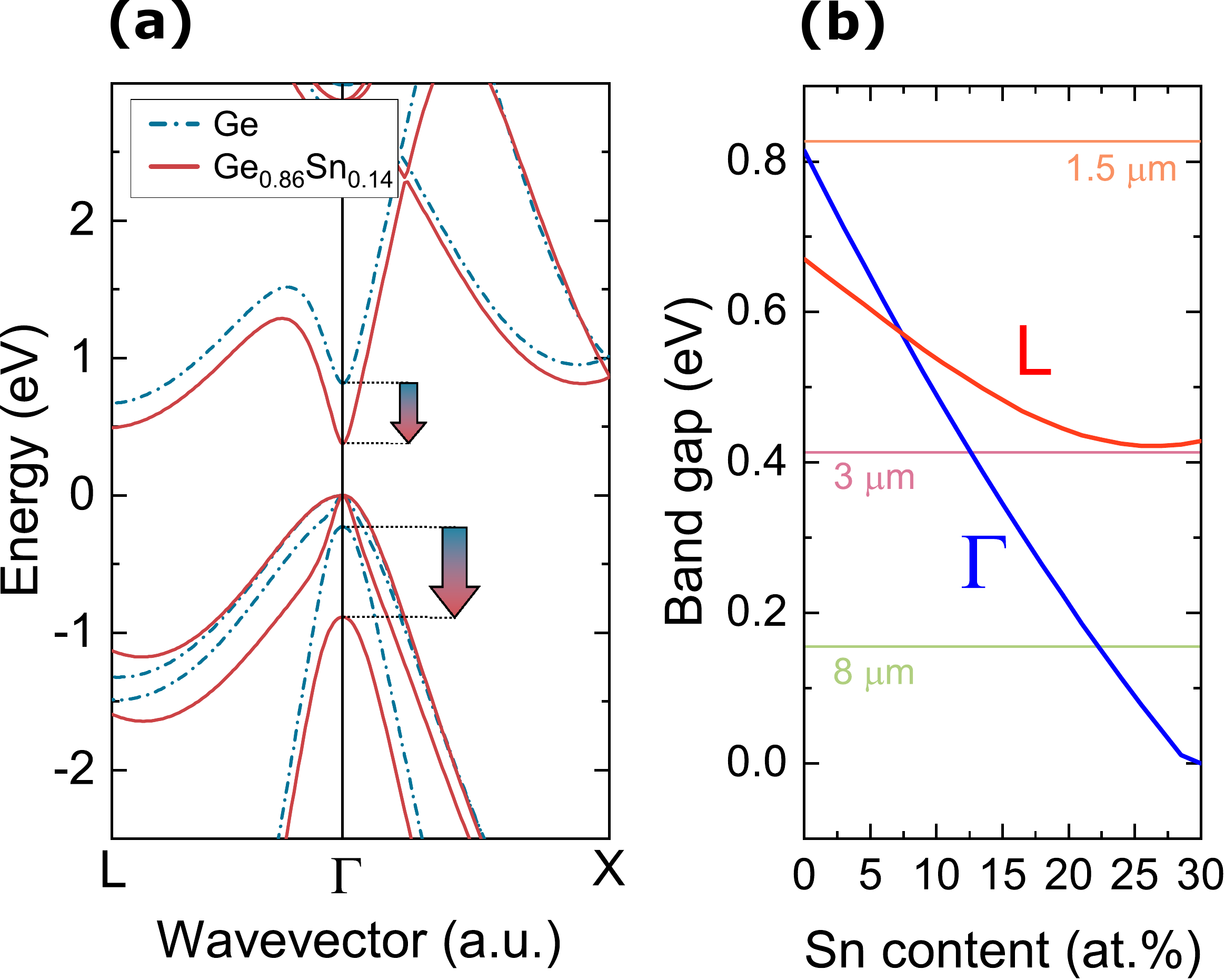}
    \caption{(a) Band structure of Ge and $Ge_{0.86}Sn_{0.14}$. (b) Band gap of Ge$_{1-x}$Sn$_{x}$ as a function of Sn content at the L and $\Gamma$ valleys.}
    \label{fig:bandstruct}
\end{figure}

 \begin{table*}[t]
  \centering
  
  \caption{$\mathrm{Ge_{1-x}Sn_x}$ parameters for the 30-band k$\cdot$p model.}
	\resizebox{\textwidth}{!}{%
	
	{\renewcommand{\arraystretch}{1.5}
  \begin{tabular}{cccccccc}
    \hline \hline $(\mathrm{eV})$ & $\mathrm{Ge_{1-x}Sn_x}$  & $(\mathrm{eV})$ & $\mathrm{Ge_{1-x}Sn_x}$ & $(\mathrm{eV}. \mathrm{nm})$ & $\mathrm{Ge_{1-x}Sn_x}$  & $(\mathrm{eV}. \mathrm{nm})$ & $\mathrm{Ge_{1-x}Sn_x}$  \\
    \hline $\Gamma_1^l$ & $-12.2519+1.4249 x$ &  $\Gamma_{1}^u$ & $8.2064-2.7334 x$  & $P_{1}$ & $0.8421+0.3350 x-1.6385 x^{2}$  & $Q_{2}$ & $-0.5334+3.4420 x+2.4647 x^{2}$  \\
$\Delta_{25'}^l$ & $0.2247+5.3808 x-4.9535 x^{2}$  & $\Gamma_{12'}$ & $8.5786-0.9856 x$  & $P_{2}$ & $0.1781-2.1954 x+1.9328 x^{2}$  & $R_{1}$ & $0.3757+0.0340 x+0.0089 x^{2}$  \\
$\Gamma_{2'}^l$ & $0.8140-3.4667 x+2.2767 x^{2}$  & $\Gamma_{25'}^u$ & $13.4041-4.8581 x$ & $P_{3}$ & $-0.0734+1.3200 x-1.2452 x^{2}$  & $R_{2}$ & $0.6820-1.1743 x$  \\
$\Gamma_{15}$ & $2.990-0.796 x$  & $\Delta_{25'}^u$ & $0.0793-0.0333 x$  & $P_{4}$ & $1.0543+0.2174 x-0.4220 x^{2}$  & $T_{1}$ & $0.7994-0.0565 x-0.0441 x^{2}$  \\
$\Delta_{15}$ & $0.2520+0.193 x$  & $\Gamma_{2'}^u$ & $17.0426-5.5226 x$  & $Q_{1}$ & $0.8114-0.1332 x-0.0055 x^{2}$ & $T_{2}$ & $-0.0384+0.3675 x$ \\
$\Gamma_{25'}^l$ & $0.00$  & $\Delta_{\Gamma_{25'}^l, \Gamma_{25'}^u}$ & $0.22+0.336 x$ &  & &  &  \\
\noalign{\vskip 2mm} \hline \hline 
  \end{tabular}}%
  }
  
  \label{tab:1}
\end{table*}

The band structure of Ge and Ge$_{0.86}$Sn$_{0.14}$, calculated using the 30-band k$\cdot$p model, are displayed in Fig.~\ref{fig:bandstruct} a). It is noticeable that the incorporation of Sn in Ge induces significant changes such as the lowering of the split-off band and of the first conduction band at the L point and $\Gamma$ point. In Fig.~\ref{fig:bandstruct} b), the bandgap at the L and  $\Gamma$ valleys are plotted as a function of the Sn content. The extended short-wave infrared (1.5 - 3$\mu$m) and mid-wave infrared (3 - 8$\mu$m) regions are fully covered by the direct gap. Note that the bandgap of bulk Ge$_{1-x}$Sn$_{x}$ becomes direct at around 8\% of Sn, which is expected to have a strong impact on the dynamics of carriers generated by the optical injection.

The evaluation of the optical response tensors is carried out following a semiclassical framework within k$\cdot$p theory, in the independent-particle approximation. We neglect the indirect absorption processes since they require phonon mediation and are thus much weaker than direct processes. The details of the derivation are described elsewhere \cite{rioux2012optical}. The calculated tensor components take the following basic form:
 
 \begin{equation}
G(\omega)=\sum_{c, v} \int \frac{d^{3} k}{8 \pi^{3}} g_{c v}(\mathbf{k}) \delta\left[\omega-\omega_{c v}(\mathbf{k})\right]{ }.
\end{equation}
The integral over the Brillouin Zone (BZ) is performed by employing a linear tetrahedron method \citep{blochl1994improved}. A typical challenge for these types of calculations is that reaching convergence can be highly demanding in terms of computation resources. By exploiting the point-group symmetry of the Ge$_{1-x}$Sn$_{x}$ crystal, the space to be sampled is reduced to an irreducible "wedge" of the BZ. In order to accurately reproduce all the features of the response tensors, the region close to the $\Gamma$ point requires a finer mesh as compared to the outer regions. Therefore, a non-uniform mesh was introduced which improves greatly the efficiency of the k-point sampling. Finally, the calculations benefit from the parallel computation by regrouping the operations at k-points in batches and distributing them to separate processor cores for simultaneous calculations.


\UseRawInputEncoding
\section{Results and Discussion}
\label{sec:Results}

\subsection{Carrier and spin injection}

In the following, the calculations of carrier and spin injection in Ge$_{1-x}$Sn$_{x}$  are described for an incident monochromatic field of a frequency $\omega$:
\begin{equation}
\mathbf{E}(t)=\mathbf{E}(\omega) e^{-i \omega t}+\text { c.c. }.
\end{equation}
The carrier and spin optical injection rates are given by: 
\begin{equation}
\dot{n}=\dot{n}_{1}(\omega)+\dot{n}_{2}(\omega){ },
\end{equation}
\begin{equation}
\dot{\mathbf{S}}=\dot{\mathbf{S}}_{1}(\omega)+\dot{\mathbf{S}}_{2}(\omega){ },
\end{equation}
where subscripts 1 and 2 are associated with the first and second-order absorption processes, respectively. The results for both processes are presented in the following two sections.

\subsubsection{One-photon absorption}

When considering only the one-photon transition amplitude, the carrier injection $\dot{n}_{1}$ is defined as:
 \begin{equation}
\dot{n}_{1}(\omega)=\xi_{1}^{a b}(\omega) E^{a *}(\omega) E^{b}(\omega){ },
\end{equation}
with the response tensor
\begin{equation}
\xi_{1}^{a b}(\omega)=\frac{2 \pi e^{2}}{\hbar^{2} \omega^{2}} \sum_{c, v} \int \frac{d^{3} k}{8 \pi^{3}} v_{c v}^{a *}(\mathbf{k}) v_{c v}^{b}(\mathbf{k}) \delta\left[\omega_{c v}(\mathbf{k})-\omega\right]{ }.
\end{equation}
Similarly, the rate of one-photon spin injection is:
\begin{equation}
\dot{S}_{1}^{a}(\omega)=\zeta_{1}^{a b c}(\omega) E^{b}(\omega) E^{c}(\omega){ }.
\label{Eq:One-ph_spin}
\end{equation}
In this case, a multiple-scale approach is employed to consider the coherence between nearly degenerate excited states \cite{nastos2007full}. The pseudotensor electron contribution $\zeta_{\substack{{1;e}}}^{a b c}$ and hole contribution $\zeta_{\substack{{1;h} }}^{a b c}$ are given by:
\begin{align}
    \zeta_{\substack{{1;e} \\ (h)}}^{a b c}(\omega)= & (-) \frac{\pi e^{2}}{\hbar^{2} \omega^{2}}\!\! \sum_{\substack{c, c^{\prime}, v \\\left(c, v, v^{\prime}\right)}}^{\prime} \!\! \int \frac{\mathrm{d}^{3} k}{8 \pi^{3}} S_{\substack{cc^{\prime} \\\left(v^{\prime} v\right)}}^{a}\!\!(\mathbf{k}) v_{c v}^{b *}(\mathbf{k}) v_{\substack{c^{\prime} v \nonumber \\\left(cv^{\prime}\right)}}^{c}\! \!(\mathbf{k}) \\ &\!\times\!\biggl(\delta\left[\omega_{c v}(\mathbf{k})-\omega\right]+\delta\biggl[\omega_{\substack{c^{\prime} v \\\left(c v^{\prime}\right)}}\!\!(\mathbf{k})-\omega\biggr]\biggr){ }.
\end{align}
As explained in Section II, the model considers a diamond crystal structure. Here, a single nonzero independent component exists for the carrier injection response tensor $\xi_{1}^{a b}$ and the spin injection pseudotensor $\zeta_{\substack{{1}}}^{a b c}$:
\begin{equation}
\xi_{1}^{x x}=\xi_{1}^{y y}=\xi_{1}^{z z}{ },
\end{equation}
\begin{equation}
\zeta_{1}^{x y z}=\zeta_{1}^{y z x}=\zeta_{1}^{z x y}=-\zeta_{1}^{x z y}=-\zeta_{1}^{y y x z}=-\zeta_{1}^{z y x}{ }.
\end{equation}
$\xi_{1}$ is related to the susceptibility of the material as $\operatorname{Im}[\chi(\omega)]=\frac{\hbar}{2} \xi_{1}(\omega)$. Within the single-particle approximation, $\xi_{1}^{x x}$ and $\zeta_{1}^{x y z}$ are purely real and purely imaginary, respectively. 

\begin{figure}[h]
    \centering
    \includegraphics[scale=0.55]{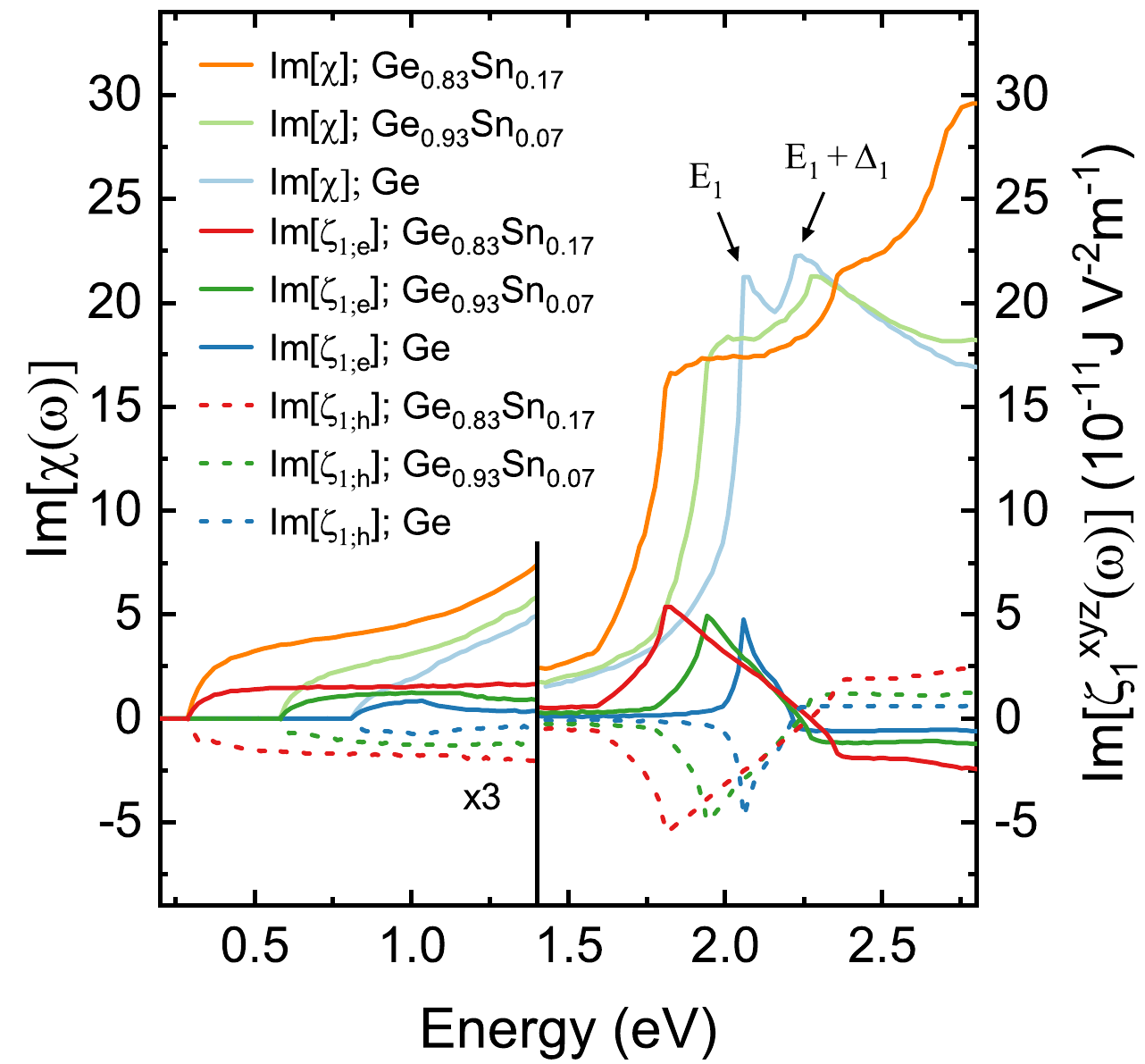}
    \caption{The first order response of Ge, Ge$_{0.93}$Sn$_{0.07}$ and Ge$_{0.83}$Sn$_{0.17}$, as a function of incident photon energy $\hbar\omega$, calculated with the 30-band k$\cdot$p model. In light colors, the imaginary part of the susceptibility $\chi(\omega)$ is presented, with the $E_1$ and $E_1+\Delta_1$ peaks identified for Ge. In darker colors, the plain (dashed) curves show the electron (hole) contribution of the spin-injection component $\zeta_{1}^{x y z}$.}
    \label{fig:OnePhInj}
\end{figure}

A selected set of the obtained results is presented in Fig.~\ref{fig:OnePhInj} for three Ge$_{1-x}$Sn$_{x}$ compositions. As the Sn content increases, the onset of absorption occurs at lower energies because of the reduction of the band gap at the $\Gamma$ point. At the same time, the $E_1$ peak of carrier injection decreases in magnitude and shifts to lower energies, while the $E_1+\Delta_1$ peak shifts to higher energies due to the stronger spin-orbit coupling. For both electron and hole contributions to spin injection, there is a broadening of the peak at the $E_1$ resonance, and the absolute value of its maximum increases slightly while shifting to lower energies.\\

In equation \eqref{Eq:One-ph_spin}, the spin injection rate is maximized by considering a circularly polarized optical field. For a $\sigma^{-}$ light propagating along the $-\hat{\mathbf{z}}$ direction, the spin polarization of the injected carriers is parallel to the z-axis. In this case, the degree of spin polarization (DSP), defined as the excess of spin-up versus spin-down polarized carriers such as $\operatorname{DSP}^{z}=\frac{n_{\uparrow}-n_{\downarrow}}{n_{\uparrow}+n_{\downarrow}}$ is calculated using the following equation, under the symmetry considerations previously discussed:
\begin{equation}
\operatorname{DSP}^{z}=\frac{2}{\hbar} \frac{\dot{S}_{1}^{z}}{\dot{n}_{1}}=\frac{2}{\hbar} \frac{\operatorname{Im}\left[\zeta_{1}^{x y z}\right]}{\xi_{1}^{x x}}{ }.
\end{equation}

 \begin{figure}[h]
    \centering
    \includegraphics[scale=0.45]{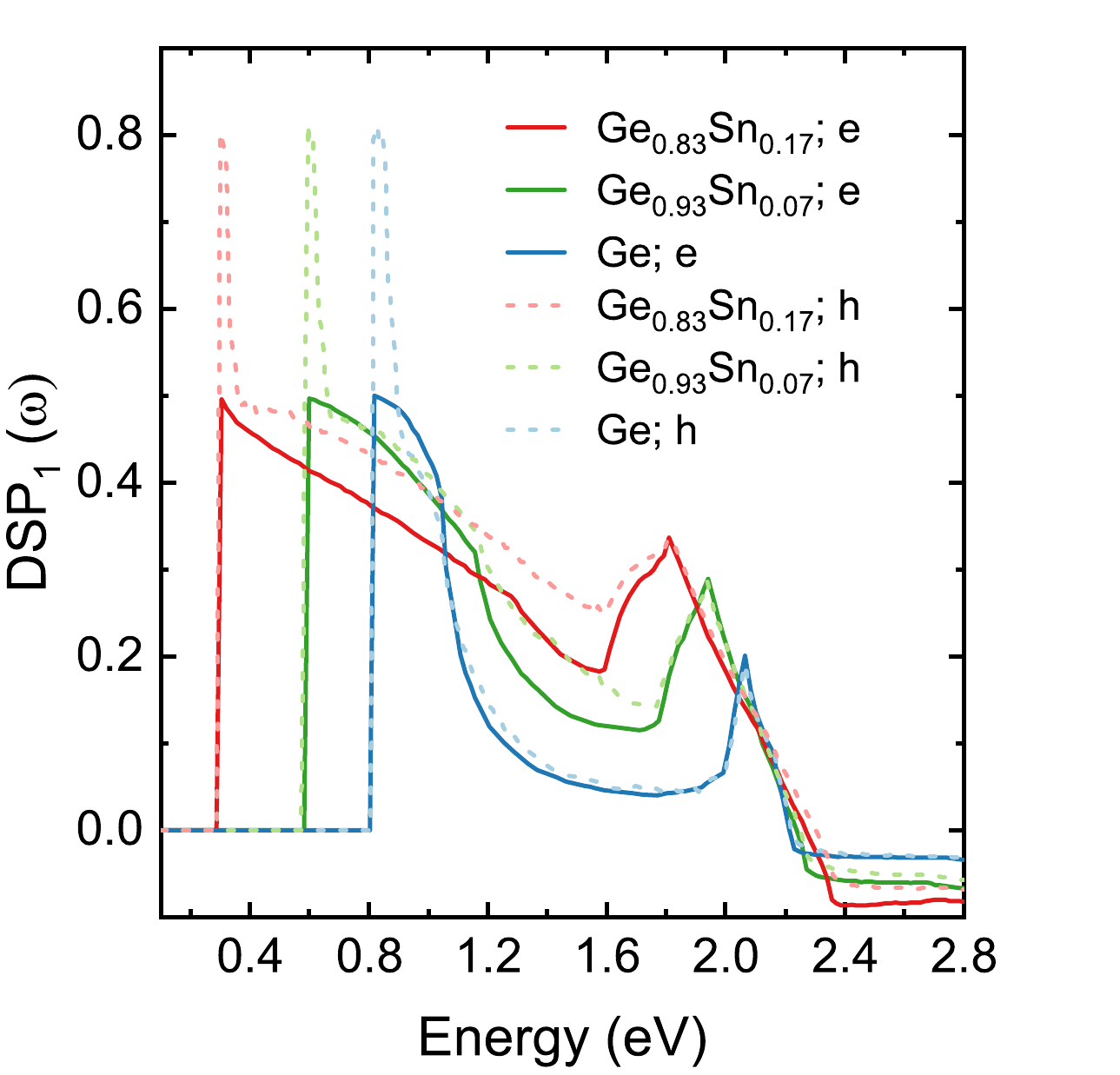}
    \caption{The degree of spin polarization of carriers optically injected in Ge, $Ge_{0.93}Sn_{0.07}$ and $Ge_{0.83}Sn_{0.17}$, by left-circularly polarized light. The plain (dashed) curves show the electron (hole) spin. The sign of the hole DSP is reversed. }
    \label{fig:OnePhDSP}
\end{figure}

The obtained results are displayed in Fig.~\ref{fig:OnePhDSP} for the same three compositions as in Fig.~\ref{fig:OnePhInj}. The reference data for Ge exhibit similar behavior to that reported earlier \cite{rioux2010optical}, and the only slight difference observed in the direct band gap value is attributed to a difference in temperature for the material parametrization employed in the two studies. The predicted existence of a DSP peak at the $E_1$ transition of Ge has been demonstrated experimentally \cite{rinaldi2014wide}. Note that the observed shift of the direct gap to lower energies as the Sn content increases is expected, as described in Fig.~\ref{fig:bandstruct} b). For the electron contribution, the DSP saturates at 50\% at the band edge, due to selection rules considerations discussed in detail in Rioux and Sipe \cite{rioux2012optical}. The decrease of the DSP triggered by the onset of spin injection from the split-off band is shifted to higher energies in Ge$_{1-x}$Sn$_x$, due to the increase of the spin-orbit coupling with the incorporation of Sn. Therefore, the DSP remains relatively high for a larger energy range, until the decline at around 1.2 eV. At the $E_1$ transition, the increase in Sn content is associated with a shift of the peak to lower energies, and the value of the DSP increases from 20\% in pure Ge to above 30\% for Ge$_{0.83}$Sn$_{0.17}$. This originates from the simultaneous drop of carrier injection and increase of spin injection observed at the $E_1$ resonance in Fig.~\ref{fig:OnePhInj}. Above 2.2 eV, the DSP takes negative values for higher Sn content.

 \subsubsection{Two-photon absorption}
 
 In semiconductors, the second-order responses to the incident field are proportional to the square of its intensity. The two-photon carrier injection is given by
 \begin{equation}
\dot{n}_{2}(\omega)=\xi_{2}^{a b c d}(\omega) E^{a *}(\omega) E^{b *}(\omega) E^{c}(\omega) E^{d}(\omega){ }.
\end{equation}
The expression for the fourth-rank tensor is:
\begin{equation}
\xi_{2}^{a b c d}(\omega)=\frac{2 \pi e^{4}}{\hbar^{4} \omega^{4}} \sum_{c, v} \int \frac{\mathrm{d}^{3} k}{8 \pi^{3}} w_{c v}^{a b *}(\mathbf{k}) w_{c v}^{c d}(\mathbf{k}) \delta\left[\omega_{c v}(\mathbf{k})-2 \omega\right]
\end{equation}

\noindent where $\boldsymbol{w}_{c v}^{a b}(\mathbf{k})$ is the symmetrized two-photon amplitude:
\begin{equation}
w_{c v}^{a b}(\mathbf{k}) \equiv \frac{1}{2} \sum_{m} \frac{v_{c m}^{a}(\mathbf{k}) v_{m v}^{b}(\mathbf{k})+v_{c m}^{b}(\mathbf{k}) v_{m v}^{a}(\mathbf{k})}{\omega_{m}(\mathbf{k})-\bar{\omega}_{c v}(\mathbf{k})}{ }.
\end{equation}
In total, by symmetry of the diamond lattice, the tensor $\xi_{2}^{a b c d}$ has 21 nonzero components. By exchanging $a \leftrightarrow b$, $c \leftrightarrow d$, $ab \leftrightarrow cd$, or performing permutations of Cartesian directions, 3 independent components are obtained: $\xi_{2}^{x x x x}$, $\xi_{2}^{x x y y}$ and $\xi_{2}^{x y x y}$. Within the independent particle approximation, they are purely real, and associated to the third-order nonlinear susceptibility $\boldsymbol{\chi}^{(3)}$ by $\operatorname{lm}\left[\chi^{(3)}(\omega ;-\omega, \omega, \omega)\right]=(\hbar / 3) \xi_{2}(\omega)$. 

\begin{figure}[h]
    \centering
    \includegraphics[scale=0.4]{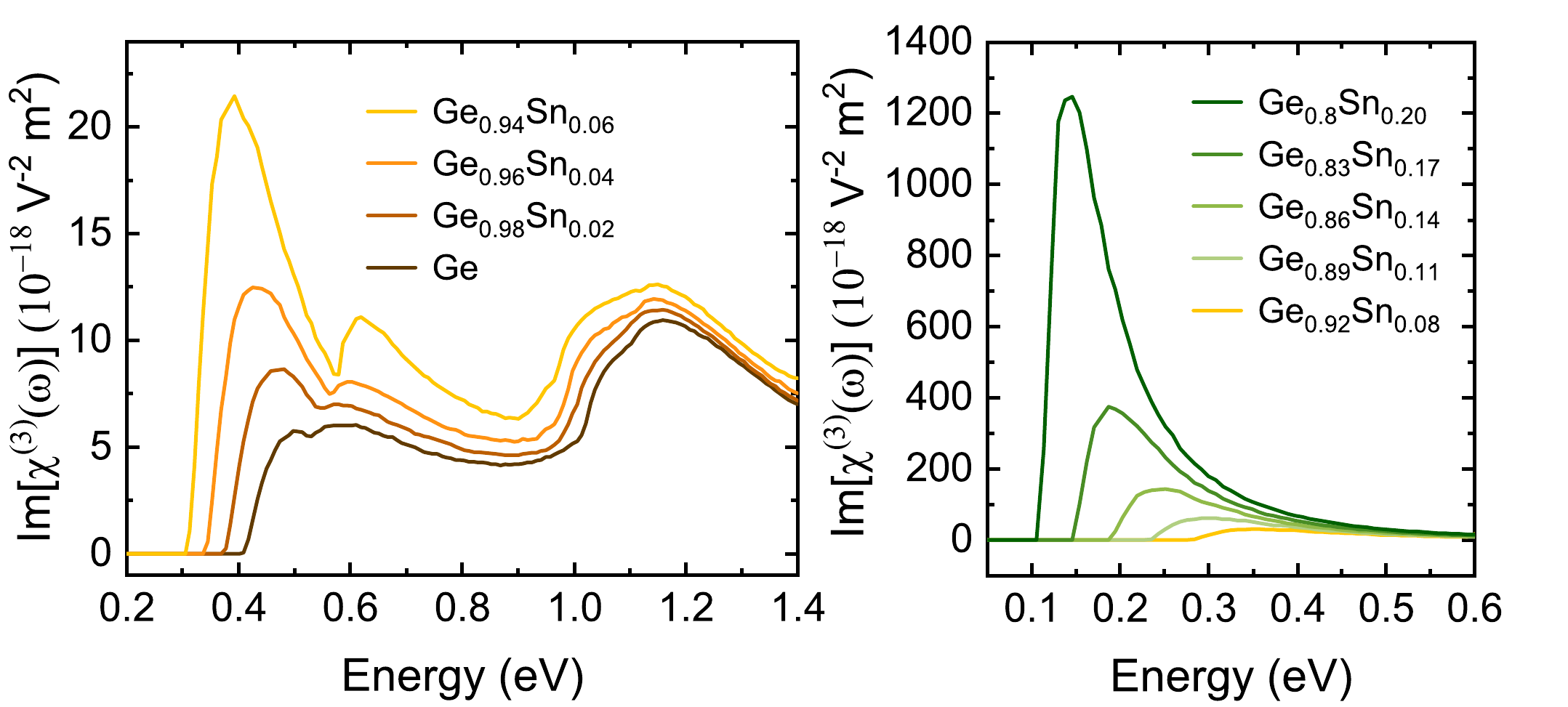}
    \caption{The component $\xi_{2}^{x x x x}$ of the two-photon absorption tensor $\xi_{2}(\omega)$ in Ge$_{1-x}$Sn$_{x}$, as a function of photon energy $\hbar\omega$ at various Sn compositions 0 to 20\%.}
    \label{fig:TwoPhCarr}
\end{figure}

The results of the calculations for the component $\xi_{2}^{x x x x}$ are displayed in Fig.~\ref{fig:TwoPhCarr}. The other components show similar features, as detailed in the Supplemental Material \citesupp{}. As the Sn content increases from 0 to 6\%, close to the band edge there is a strong augmentation of carrier injection. At approximately 0.55 eV, the onset of absorption from the split-off band generates a second regime of injection, which becomes more pronounced for higher Sn content, and shifts slightly to higher photon energies because of the higher spin-orbit coupling. Above 1 eV, there is a small rise in the response tensor. For Sn compositions from 8 to 20\%, the focus is kept on a restricted range of energies close to the band edge. A sizable increase in two-photon carrier injection, nearly exponential as a function of Sn content, is observed. This behavior is attributed to the reduction of the effective mass as Sn composition increases. \\

Analogously to carrier injection, the rate of two-photon spin injection is: 
\begin{equation}
\dot{S}_{2}^{a}(\omega)=\zeta_{2}^{a b c d e}(\omega) E^{b *}(\omega) E^{c *}(\omega) E^{d}(\omega) E^{e}(\omega){ },
\end{equation}
with the fifth-rank pseudotensor defined as:
\begin{equation}
\begin{aligned}
\zeta_{\substack{2; e \\
(h)}}^{a b c d e}(\omega)=&(-) \frac{\pi e^{4}}{\hbar^{4} \omega^{4}} \! \! \sum_{\substack{c, c{ }^{\prime}, v \\
\left(c, v, v^{\prime}\right)}}^{\prime} \! \! \int \frac{\mathrm{d}^{3} k}{8 \pi^{3}} S_{\substack{c c^{\prime} \\ \left( v^{\prime}v \right) }}^{a} \! \! (\mathbf{k}) w_{c v}^{b c *}(\mathbf{k}) w_{\substack{c^{\prime} v \\
\left(c v^{\prime}\right)}}^{d e} \! \! (\mathbf{k}) \\
& \!\times\!\left(\delta\left[\omega_{c v}(\mathbf{k})-2 \omega\right]+\delta\left[\omega_{\substack{c^{\prime} v \\
\left(c v^{\prime}\right)}}\!\!(\mathbf{k})-2 \omega\right]\right) .
\end{aligned}
\end{equation}

\noindent By symmetry, $\zeta_{2}$ has two independent components: $\zeta_{2}^{x x y x z}$ and $\zeta_{2}^{x y z z z}$. All 48 nonzero components of the pseudotensor can be retrieved by applying cyclic permutations of the Cartesian directions, exchanging $b \leftrightarrow c$, $d \leftrightarrow e$, or any combination of these. Within the single-particle approximation, $\zeta_{2}$ is purely imaginary. 
 
 \begin{figure}[h]
    \centering
    \includegraphics[scale=0.4]{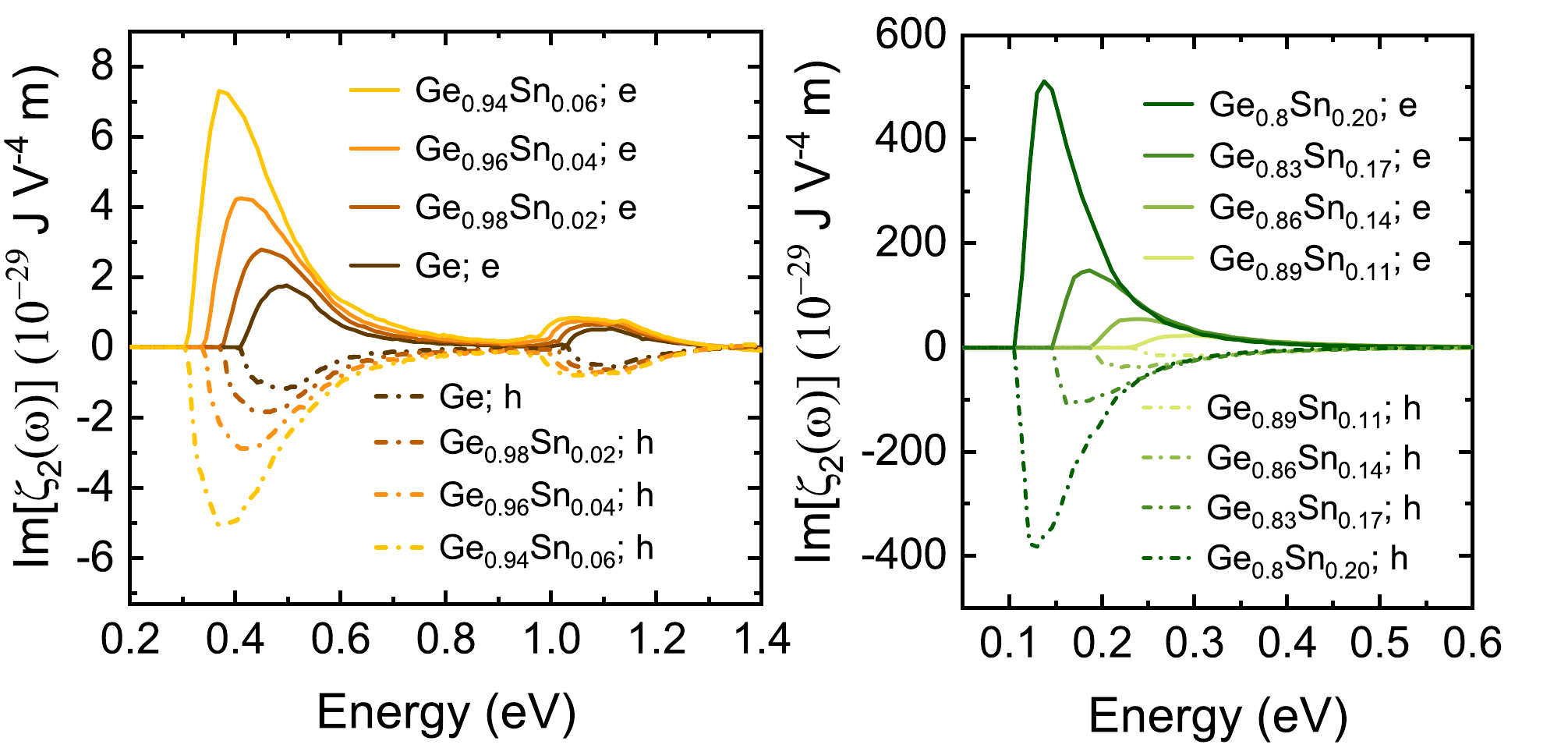}
    \caption{The component $\zeta_{2}^{x x y x z}$ of the two-photon spin injection pseudotensor $\zeta_{2}(\omega)$ in Ge$_{1-x}$Sn$_{x}$, as a function of photon energy $\hbar\omega$. The electron (hole) contribution takes positive (negative) values. The different curves show various Sn compositions of the alloy, from 0 to 20\%. }
    \label{fig:TwoPhSpin}
\end{figure}

The results for the component $\zeta_{2}^{x x y x z}$ are presented in Fig.~\ref{fig:TwoPhSpin}. The other component shows similar features and is discussed in detail in the Supplemental Material \cite{Note1}. For Ge$_{1-x}$Sn$_{x}$ alloys with Sn content up to 6\%, at the onset of absorption the spin injection increases significantly for both electrons and holes. However, the absolute value of the magnitude of this first peak is relatively higher for the electron contribution. At the $E_1$ transition around 1.1 eV, an asymmetric broadening of the peak is observed as Sn content increases. In GeSn alloys with even higher Sn compositions reaching up to 20\%, the spin injection becomes increasingly strong and similar to the exponential increase of carrier injection discussed above. Similarly, this drastic increase originates from the reduction in the effective mass of electrons and holes observed when the Sn content increases in the alloy. \\

From the two-photon carrier and spin injection, the two-photon DSP can be evaluated. The anisotropy described previously leads to a dependence of the two-photon DSP on the orientation of the beam. For $\boldsymbol{\sigma}^{-}$ light incident along <001>, the degree of spin polarization is given by:
\begin{equation}
\mathrm{DSP}_{2}^{\langle 001\rangle}=\frac{(\hbar / 2)^{-1} 2 \operatorname{Im}\left[\zeta_{2}^{x y z z z}(\omega)\right]}{\xi_{2}^{x y x y}(\omega)-\frac{1}{2} \xi_{2}^{x x y y}(\omega)+\frac{1}{2} \xi_{2}^{x x x x}(\omega)}{ }.
\end{equation}

\begin{figure}[h]
    \centering
    \includegraphics[scale=0.4]{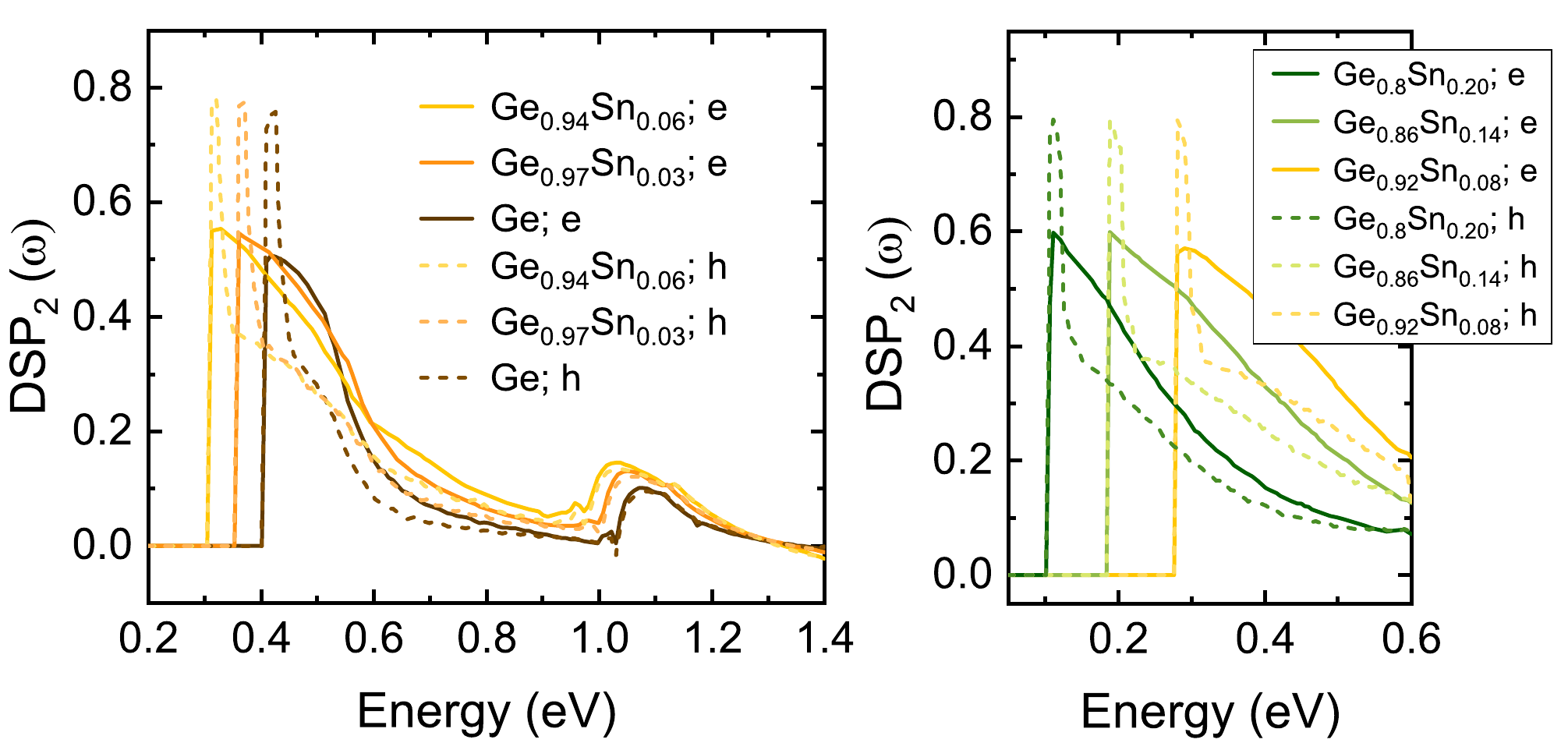}
    \caption{The degree of spin polarization of carriers optically injected by two-photon absorption in Ge$_{1-x}$Sn$_{x}$, for left-circularly polarized light at <001> incidence. The plain (dashed) curves show the electron (hole) spin. The sign of the hole DSP is reversed. The different curves show various Sn compositions of the alloy, from 0 to 20\%.}
    \label{fig:TwoPhDSP}
\end{figure}

The results of the calculations are presented in Fig.~\ref{fig:TwoPhDSP}. The two-photon electron DSP shows similar features to the one-photon DSP. However, close to the band edge, in the case of Ge, the value is slightly above 50\%. As the Sn content increases, contrary to the one-photon DSP which saturates at a maximum value fixed by the selection rules, the two-photon DSP at the band edge increases to reach approximately 60\% for alloys with a Sn content above 14\%. In the case of holes, a similar small increase in absolute value is observed with the incorporation of Sn leading to two-photon DSP around -80\%. At the $E_1$ resonance, for the two types of carriers, there is a broadening of the peak, and its maximum increases slightly while shifting to lower energies.

\subsection{Coherent control}

\begin{figure*}
    \centering
    \includegraphics[width=16cm]{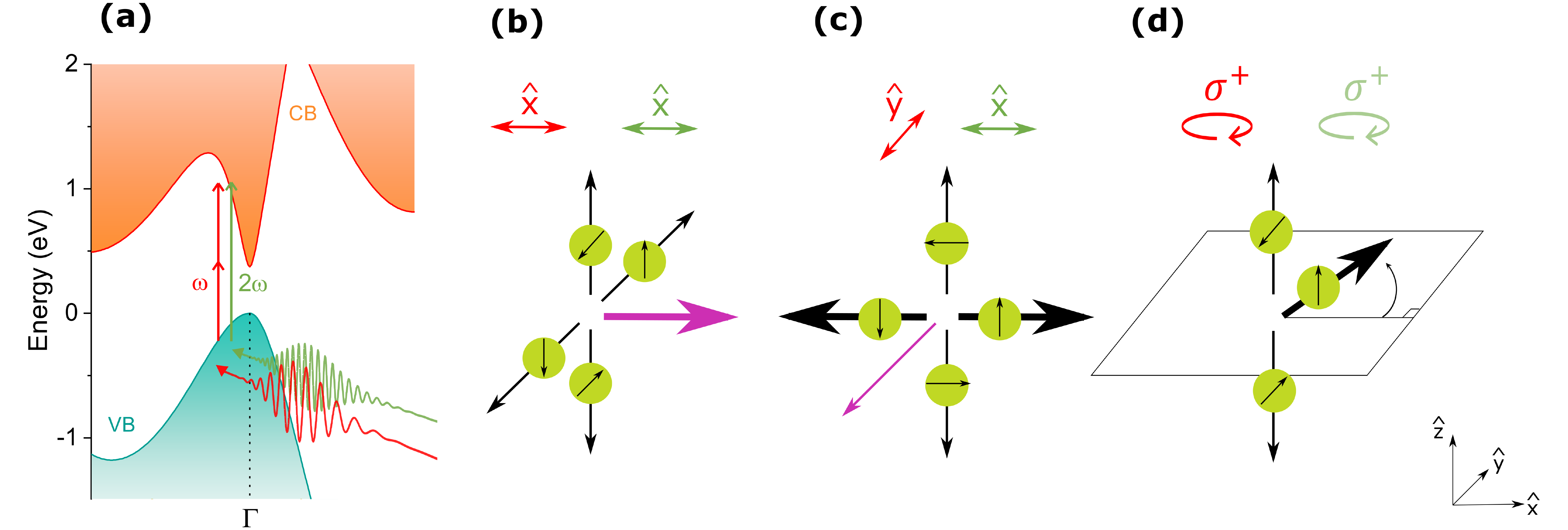}
    \caption{ \textbf{Two-color coherent control scheme.} (a) Electronic band structure of Ge$_{0.86}$Sn$_{0.14}$. An optical field at angular frequency 2$\omega$ can excite direct transitions from the valence band to the conduction band, while another field at $\omega$ can excite two-photon transitions. The polarization of the two fields determines the motion and spin of the injected carriers. (b)-(d) Carrier motion and spin polarization in the two-color coherent control scheme, for light beams incident on <001> with (b) co-linear, (c) cross-linear, and (d) co-circular polarization. The red (green) arrows indicate the polarization state of the beam at angular frequency $\omega$ (2$\omega$). Purple arrows indicate charge currents, black arrows indicate spin-polarized currents, and counter-propagating black arrows indicate pure spin currents.}
    \label{fig:Configurations}
\end{figure*} 

In this section, a bichromatic field of frequencies $\omega$ and $2\omega$ is considered:
\begin{equation}
\mathbf{E}(t)=\mathbf{E}(\omega) e^{-i \omega t}+\mathbf{E}(2 \omega) e^{-2 i \omega t}+\text { c.c. }.
\end{equation}
 The transition energy from two-photon absorption of the beam with frequency $\omega$ matches that of the one-photon absorption of the beam with frequency $2\omega$. The interference between both pathways leads to the injection of charge and spin currents in the semiconductor.
\subsubsection{Charge and spin currents}

The charge current has an injection rate given by:
\begin{equation}
\dot{J}_{I}^{a}=\eta_{I}^{a b c d}(\omega) E^{b *}(\omega) E^{c *}(\omega) E^{d}(2 \omega)+\text { c.c. },
\end{equation}
with the fourth-rank tensor $\eta_{I}$:
\begin{align}
\eta_{I}^{a b c d}(\omega)=&(-) \frac{i \pi e^{4}}{\hbar^{3} \omega^{3}} \sum_{c, v} \int \frac{d^{3} k}{8 \pi^{3}} v_{\substack{c c \\(v v)}}^{a}(\mathbf{k}) w_{c v}^{b c *}(\mathbf{k}) v_{c v}^{d}(\mathbf{k}) \nonumber \\
& \times \delta\left[\omega_{c v}(\mathbf{k})-2 \omega\right] .
\end{align}

\noindent In the diamond structure symmetry, there are 21 nonzero components related to three independent components $\eta_{I}^{x x x x}$, $\eta_{I}^{x x y y}$ and $\eta_{I}^{x y y x}$, by cyclic permutations of the indices and the exchanges $b \leftrightarrow c$, $ab \leftrightarrow cd$. 

Similar to the calculation of charge current, the spin-current injection is evaluated with the following expression:
\begin{equation}
\dot{K}_{I}^{a b}=\mu_{I}^{a b c d e}(\omega) E^{c *}(\omega) E^{d *}(\omega) E^{e}(2 \omega)+\text { c.c. }.
\end{equation}
$\mu_{I}$ is a fifth-rank pseudotensor obtained by employing the multiple-scale approach:
\begin{equation}
\begin{aligned}
\mu_{I}^{a b c d e}(\omega)\! = & (-) \frac{i \pi e^{3}}{2 \hbar^{3} \omega^{3}} \! \! \! \! \sum_{\substack{c, c{ }^{\prime}, v \\
\left(c, v, v^{\prime}\right)}} \! \! \! \int \! \frac{d^{3} k}{8 \pi^{3}} K_{\substack{c, c{ }^{\prime} \\
\left(v^{\prime}, v\right)}}^{a b} \! \! \! (\mathbf{k}) w_{c v}^{c d *}(\mathbf{k}) v_{\substack{c{ }^{\prime}, v \\
\left(c, v^{\prime}\right)}}^{e} \! \! (\mathbf{k})\\
&\times\left(\delta\left[\omega_{c v}(\mathbf{k})-2 \omega\right]+\delta\left[\omega{ }_{c^{\prime} v}(\mathbf{k})-2 \omega\right]\right),\\
\end{aligned}
\end{equation}                     

\noindent where $K_{m n}^{a b}(\mathbf{k})$ is the spin-current matrix element between bands m and n at a wavevector k:
\begin{equation}
\left\langle m \mathbf{k}\left|v^{a} S^{b}\right| n \mathbf{k}^{\prime}\right\rangle=K_{m n}^{a b} \delta\left(\mathbf{k}-\mathbf{k}^{\prime}\right){ }.
\end{equation}

\subsubsection{Configurations and composition-related behavior}

To perform a coherent control experiment in the case of a bichromatic field of frequencies $\omega$ and 2$\omega$, there are three different typical optical schemes, as shown and described in Fig.~\ref{fig:Configurations}. \\

\paragraph{Colinearly polarized beams.}

Under this configuration, the two optical fields are colinearly polarized along the x direction. A strong charge current is injected in the axis of polarization, following the equation below, while weak spin currents are injected in the y-z plane. 

\begin{equation}
\dot{\mathbf{J}}_{I}=2 \operatorname{Im}\left[\eta_{I}^{x x x x}\right] \hat{\mathbf{x}} E_{\omega}^{2} E_{2 \omega} \sin (\Delta \phi),
\end{equation}
with the phase-matching parameter $\Delta \phi \equiv 2 \phi_{\omega}-\phi_{2 \omega}$. 

\begin{figure}[h]
    \centering
    \includegraphics[scale=0.4]{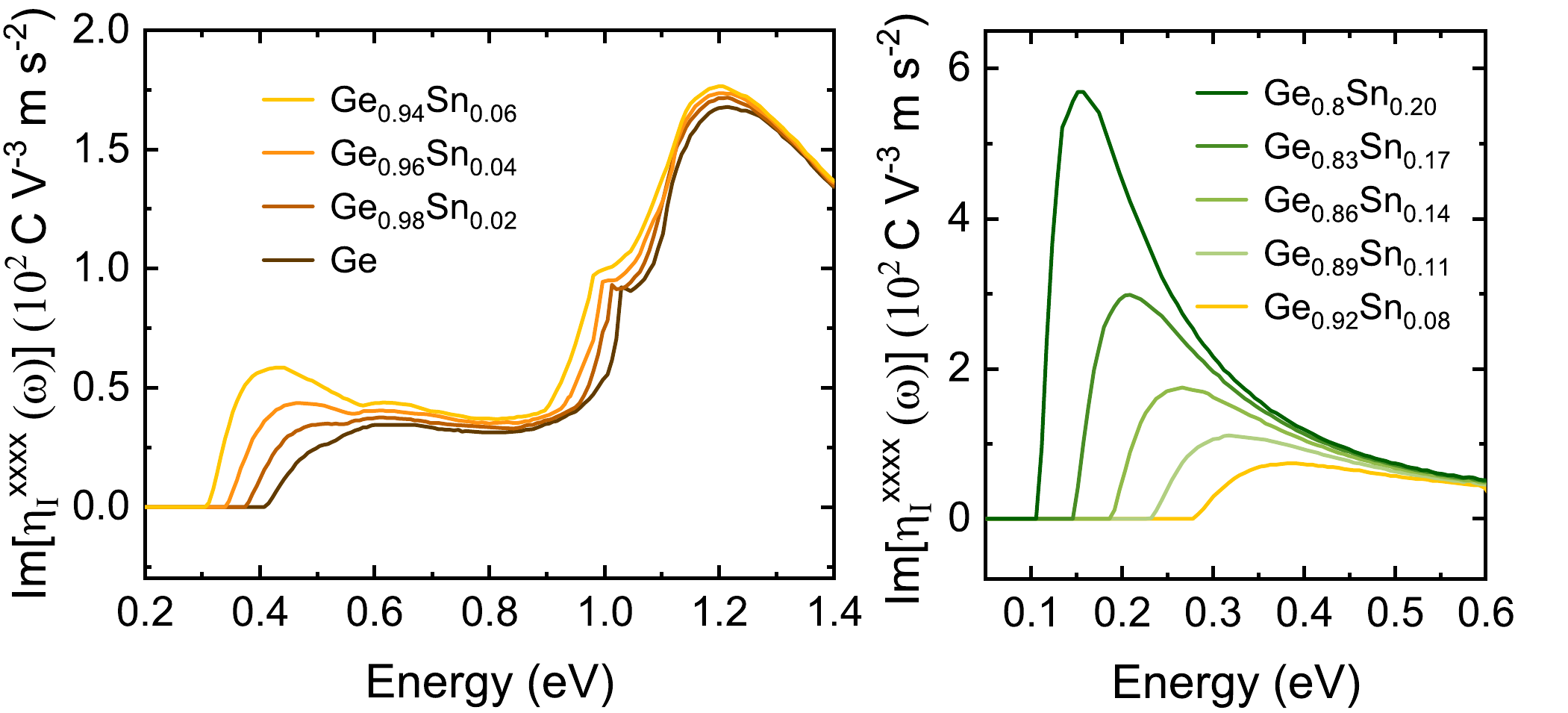}
    \caption{The electron contribution to the component $\eta_{I}^{x x x x}$ of the charge current injection tensor $\eta_{I}$ in Ge$_{1-x}$Sn$_{x}$, as a function of photon energy $\hbar\omega$. The different curves show the evolution at various Sn compositions from 0 to 20\%.}
    \label{fig:chargecurrent}
\end{figure}

The results for $\eta_{I}^{x x x x}$ as a function of the energy of the incident field are presented in Fig.~\ref{fig:chargecurrent}. The focus here is on the electron contribution since holes in Ge have a significantly shorter spin lifetime than electrons \cite{pezzoli2012optical}. In Ge, the response tensor is characterized by a strong peak at the $E_1$ resonance around 1.2 eV, compared to the smaller magnitude at lower energies. As the Sn content of the material increases, the charge current injection becomes more significant close to the band edge. Around 0.6 eV, the split-off transitions limit the injection to a stable value. Considering alloys with a Sn composition between 8 and 20\%, an exponential increase in spin current injected is observed as a function of Sn content. With the reduction of the band gap, the absorption is triggered at lower energies, and by tuning the composition, the full mid-infrared range becomes accessible. \\

\paragraph{Cross-linearly polarized beams.}

In this scheme, the optical field of frequency $2\omega$ is polarized along y, and the field of frequency $\omega$ is cross-linearly polarized along x. The main feature of this configuration is the strong pure spin-current injected along the x-axis, which is calculated with the following equation. Two other currents of relatively small magnitude are also generated: a charge current along y (expression not shown here) and a PSC along z .

\begin{equation}
\dot{K}_{I}^{a b}=2\left(\mu_{I}^{x y y y z} \hat{\mathbf{z}}^{a} \hat{\mathbf{x}}^{b}-\mu_{I}^{x y x x z} \hat{\mathbf{x}}^{a} \hat{\mathbf{z}}^{b}\right) E_{\omega}^{2} E_{2 \omega} \cos (\Delta \phi){ }.
\end{equation}

\begin{figure}[h]
    \centering
    \includegraphics[scale=0.4]{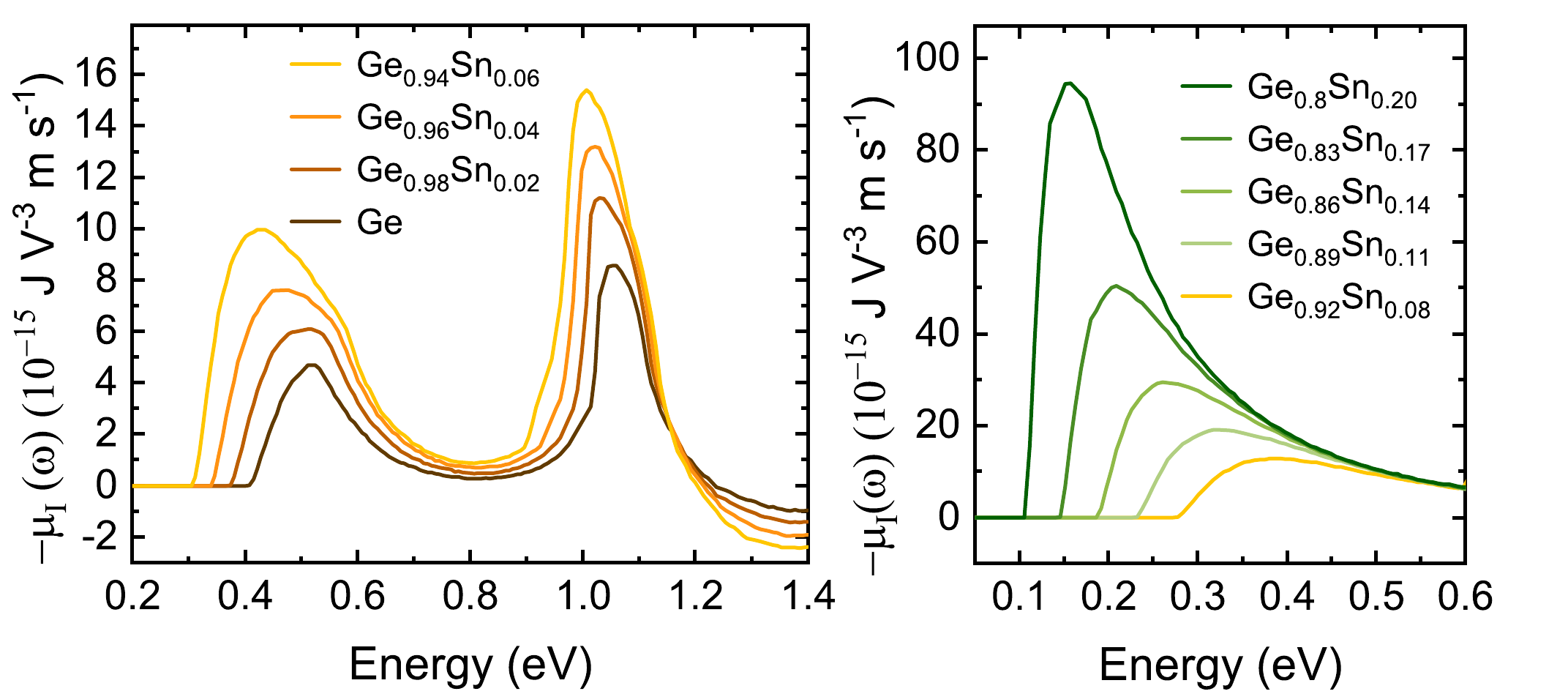}
    \caption{The electron contribution to the component $\mu_{I}^{x y x x z}$ of the spin current injection pseudotensor $\mu_{I}$ in Ge$_{1-x}$Sn$_{x}$ as a function of photon energy $\hbar\omega$. The different curves show various Sn compositions of the alloy, from 0 to 20\%.}
    \label{fig:spincurrent}
\end{figure}

The results for the pseudotensor component $\mu_{I}^{x y x x z}$ in Ge$_{1-x}$Sn$_{x}$ are presented in Fig.~\ref{fig:spincurrent}. When Sn is incorporated in Ge, there is a general increase of spin current injection for a relatively large range of energies. Compared to the charge current injection discussed previously, in the case of spin-current there is an equivalent growth of the response tensor at the $E_1$ transition and at the onset of absorption. When Sn incorporation is further enhanced to a composition of up to 20\%, there is a massive increase of spin current injection close to the band edge, which corresponds to energies in the mid-infrared region. \\

\paragraph{Circularly polarized beams.}

The last configuration employs circular polarization for the two beams. A strong spin-polarized current in the x-y plane is generated, with a direction depending on the phase difference between the two beams, as described by the following equations for the charge and spin currents. A weak PSC is also injected along the z-axis.

\begin{equation}
\dot{\mathbf{J}}_{I}=\frac{1}{\sqrt{2}} \operatorname{Im}\left[\eta_{I}^{x x x x}-\eta_{I}^{x y y x}+2 \eta_{I}^{x x y y}\right] E_{\omega}^{2} E_{2 \omega} \hat{\mathbf{m}}{ },
\end{equation}

\begin{equation}
\begin{aligned}
\dot{K}_{I}^{a b}=& \pm \frac{1}{\sqrt{2}}\left(\mu_{I}^{x y z z z}-\mu_{I}^{x y x x z}+2 \mu_{I}^{x y x z x}\right) E_{\omega}^{2} E_{2 \omega} \hat{\mathbf{m}}^{a} \hat{\mathbf{z}}^{b} \\
& \mp \frac{1}{\sqrt{2}}\left(\mu_{I}^{x y z z z}-\mu_{I}^{x y y y z}+2 \mu_{I}^{x y y z y}\right) E_{\omega}^{2} E_{2 \omega} \hat{\mathbf{z}}^{a} \hat{\mathbf{m}}^{b}{ },
\end{aligned}
\end{equation}

\noindent where $\hat{\mathbf{m}}=\hat{\mathbf{x}} \sin (\Delta \phi) \pm \hat{\mathbf{y}} \cos (\Delta \phi)$. The results for the charge and spin current injection are described in the Supplemental Material \cite{Note1}.


\UseRawInputEncoding
\section{Conclusion}
\label{sec:Conclusion}

This study demonstrates that Ge$_{1-x}$Sn$_{x}$ semiconductors can be used to modulate spin injection and coherent control in the mid-infrared range. A detailed investigation of the optical injection with the two-color quantum interference scheme is reported as a function of Sn content. One-photon and two-photon absorption processes were analyzed, and the degrees of spin polarization are extracted and their behavior is elucidated for alloys starting from pure Ge up to a Sn content of 20\%. By means of the full-zone 30-band k$\cdot$p theory, high energy features of the band structure such as the $E_1$ transition were included in the study. In addition to the expected lowering of the direct gap, the incorporation of Sn in Ge was found to induce a significant increase of the one-photon DSP at the $E_1$ resonance. In the case of two-photon injection and coherent control, the magnitude of the response tensors close to the band edge surges exponentially as a function of the Sn content. This strong increase can be attributed to the Sn incorporation-induced decrease in the carrier effective masses. It seems that this trend is also valid at the $E_1$ resonance for pure spin current injection, at least for low-Sn compositions. At the band edge, the two-photon DSP for electrons exceeds the value of 53\% in Ge, to attain 60 \% for a Sn content higher than 14 \%. A higher DSP absolute value is observed for holes reaching -80 \%. It is clear that the incorporation of Sn in Ge can be used to tune the wavelength range where a significant injection of charge and spin current can be achieved, thus providing additional degrees of freedom for the quantum coherent manipulation in the molecular fingerprint region.

\bigskip
\noindent {\textbf{ACKNOWLEDGEMENTS}}.
O.M. acknowledges support from NSERC Canada (Discovery, SPG, and CRD Grants), Canada Research Chairs, Canada Foundation for Innovation, Mitacs, PRIMA Qu\'ebec, and Defense Canada (Innovation for Defense Excellence and Security, IDEaS).\\

\bigskip

\bibliography{main.bib} 
\bibliographystyle{apsrev4-2} 

\end{document}